\documentclass[twocolumn,twocolappendix]{aastex631}

\newcommand{\cMpch}{{\rm cMpc}\,h^{-1}}
\newcommand{\hcMpc}{h\,{\rm cMpc^{-1}}}
\newcommand{\kms}{{\rm km\,s^{-1}}}
\newcommand{\msun}{M_\odot}
\shorttitle{BTPS: High-Redshift Galaxies}
\shortauthors{Hirano and Yoshida}

\begin{document}

\title{Early Structure Formation from Primordial Density Fluctuations \\with a Blue, Tilted Power Spectrum: High-Redshift Galaxies}

\correspondingauthor{Shingo Hirano}
\email{shingo-hirano@kanagawa-u.ac.jp}

\author[0000-0002-4317-767X]{Shingo Hirano}
\affiliation{Department of Astronomy, School of Science, The University of Tokyo, 7-3-1 Hongo, Bunkyo, Tokyo 113-0033, Japan}
\affiliation{Department of Applied Physics, Faculty of Engineering, Kanagawa University, Kanagawa 221-0802, Japan}

\author[0000-0001-7925-238X]{Naoki Yoshida}
\affiliation{Department of Physics, School of Science, The University of Tokyo, 7-3-1 Hongo, Bunkyo, Tokyo 113-0033, Japan}
\affiliation{Research Center for the Early Universe, School of Science, The University of Tokyo, 7-3-1 Hongo, Bunkyo, Tokyo 113-0033, Japan}
\affiliation{Kavli Institute for the Physics and Mathematics of the Universe (WPI), UT Institutes for Advanced Study, The University of Tokyo, Kashiwa, Chiba 277-8583, Japan}

\begin{abstract}
Recent observations by the James Webb Space Telescope (JWST) discovered unexpectedly abundant luminous galaxies at high redshift, posing possibly a severe challenge to popular galaxy formation models.
We study early structure formation in a cosmological model with a blue, tilted power spectrum (BTPS) given by $P(k) \propto k^{m_{\rm s}}$ with $m_{\rm s} > 1$ at small length scales.
We run a set of cosmological $N$-body simulations and derive the abundance of dark matter halos and galaxies under simplified assumptions on star formation efficiency. 
The enhanced small-scale power allows rapid nonlinear structure formation at $z>7$, and galaxies with stellar mass exceeding $10^{10}\,\msun$ can be formed by $z=9$.
Because of frequent mergers, the structure of galaxies and galaxy groups appears clumpy. 
The BTPS model reproduces the observed stellar mass density at $z=7-9$, and thus eases the claimed tension between galaxy formation theory and recent JWST observations. 
The large-scale structure of the present-day Universe is largely unaffected by the modification of the small-scale power spectrum.
We conduct a systematic study by varying the slope of the small-scale power spectrum to derive constraints on the BTPS model from a set of observations of high-redshift galaxies.
\end{abstract}

\keywords{
Cosmology (343) --- 
Dark matter (353) ---
Early universe (435) --- 
Galaxy formation (595) --- 
Population III stars (1285)
}

\section{Introduction}

The so-called $\Lambda$ Cold Dark Matter ($\Lambda$CDM) cosmology successfully reproduces a broad range of observations of the large-scale structure of the Universe, and thus has been established as the standard cosmological model.
A vital element of the standard model is the primordial density fluctuations generated in the very early universe with a nearly scale-independent power spectrum.
While the large-scale density fluctuations have been observationally probed to $k\sim1\,{\rm Mpc}^{-1}$, the amplitude and the shape of the power spectrum on smaller, (sub-)galactic length scales are poorly constrained \citep[e.g.,][]{Hlozek2012, Bullock2017}.
Hence, theoretical studies on galaxy formation often rely on significant extrapolation of the assumed scale-invariant primordial power spectrum (PPS).

Various possibilities have been proposed from the physics of the early universe that posit deviations from scale-invariance.
Blue, tilted, or enhanced power spectra can arise in beyond-standard cosmological models \citep[e.g.,][]{Covy1999, Martin2001, Gong2011, Inman2023}, and yield several interesting cosmological and astrophysical consequences \citep[e.g.,][]{Clesse2015, Germani2017}.
Earlier in \cite{Hirano2015blue}, we studied early structure formation in models with a blue, tilted power spectrum (BTPS).
It was shown that the enhanced small-scale density fluctuations drive the formation of nonlinear structure and Population III stars at very early epochs.

In this paper, we study the formation and abundance of the first galaxies in the BTPS model in light of recent observations by the James Webb Space Telescope (JWST).
Some galaxies (candidates) with unexpectedly high stellar masses have been discovered \citep[e.g.,][]{Finkelstein2022, Labbe2023}.
\citet{Boylan-Kolchin2023} concludes that the inferred high stellar masses of the observed galaxy candidates require an extremely high star formation efficiency far exceeding the plausible values of $\epsilon \lesssim 0.3$ suggested by popular galaxy formation models \citep{Gribel2017, Tacchella2018, Behroozi2020}.
The challenge brought by recent JWST observations motivates us to reconsider the detailed physics of galaxy formation in the early universe or the standard cosmology model.

Modification of PPS may provide a viable solution by promoting early structure formation \citep{Parashari2023, Padmanabhan&Loeb2023}.
\citet{Parashari2023} compute the cumulative comoving stellar mass density (CCSMD) by adopting a modified form of PPS similar to \citet{Hirano2015blue}.
Their model can successfully reproduce the observed CCSMD {\it without} assuming an unrealistically high star formation efficiency. 
They further argue that observations of high-redshift galaxies can provide invaluable information on small-scale density fluctuations that cannot be directly probed.
It is crucial and timely to study the formation of high-redshift galaxies by performing cosmological simulations for the BTPS model.
By calculating the halo mass functions for a wide range of models with varying the initial power spectrum, we are able to derive constraints on the PPS from the observed CCSMD. 
We also compute the ultra-violet luminosity function (UVLF) of galaxies in the BTPS model and compare them with the Hubble Space Telescope observations.

Throughout the present paper, we adopt the cosmological parameters with total matter density $\Omega_{\rm m} = 0.3153$, baryon density $\Omega_{\rm b} = 0.0493$ in units of the critical density, a Hubble constant $H_0 = 67.36\,\kms\,{\rm Mpc^{-1}}$, the root-mean-square matter fluctuation averaged over a sphere of radius $8\,h^{-1}\,{\rm Mpc}$ $\sigma_8 = 0.8111$, and primordial index $n_{\rm s} = 0.9649$ \citep{PlanckCollaboration2020}.

\section{Numerical simulations}
\label{sec:method}

We largely follow the method of our previous study \citep{Hirano2015blue} to perform cosmological $N$-body simulations for three sets of models with different PPS.

\begin{figure}[t]
\begin{center}
\includegraphics[width=1.0\columnwidth]{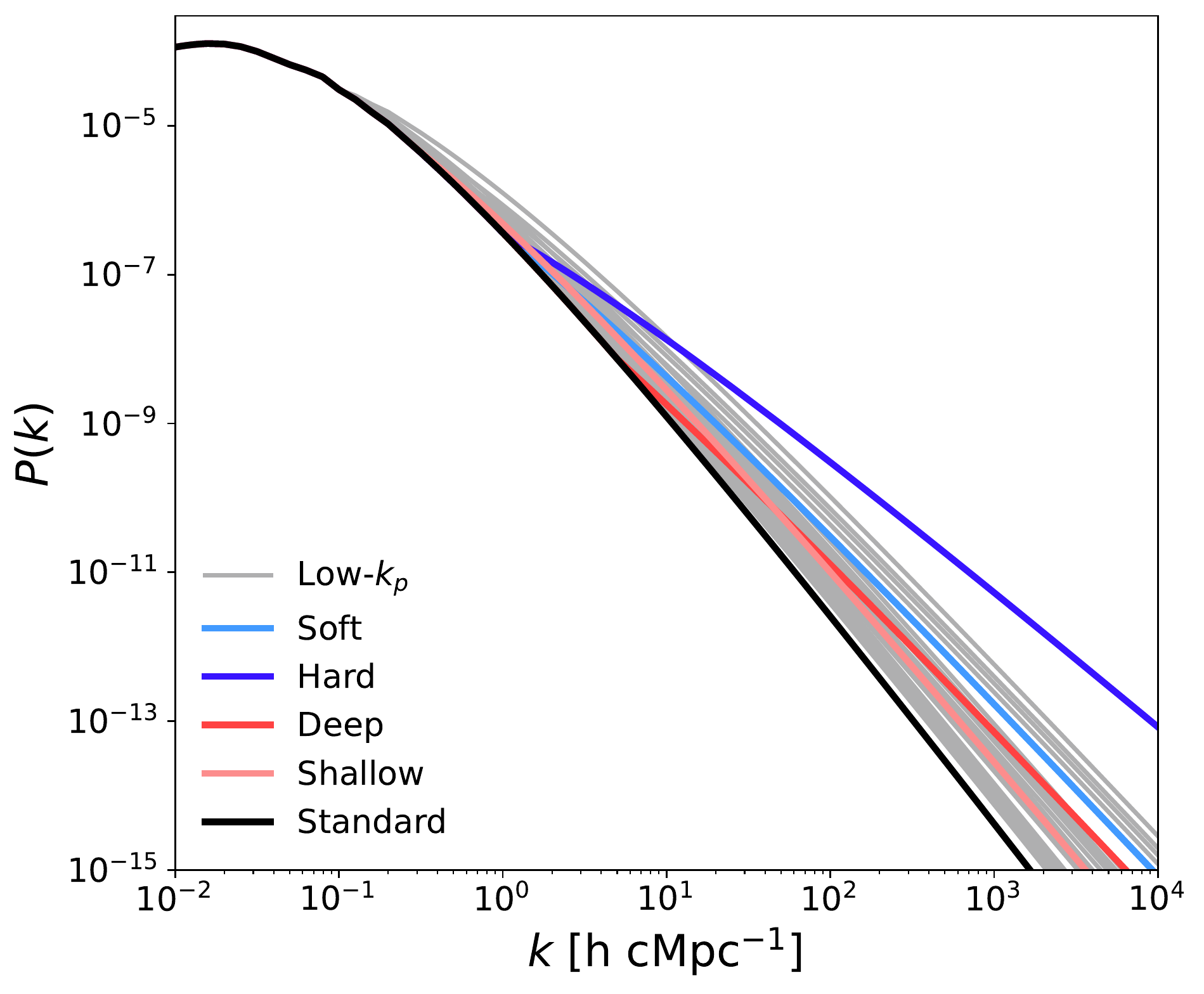}
\end{center}
\caption{
The matter power spectra at $z = 1089$ that we use to generate the cosmological initial conditions.
The black line is for the standard model, whereas the other lines show the blue-tilted models with different combinations of the pivot scale and tilt as \{$k_{\rm p}/(\hcMpc)$, $m_{\rm s}$\} = \{1.0, 1.5\} (soft), \{1.0, 2.0\} (hard), \{5.0, 1.5\} (deep), \{0.3, 1.2\} (shallow), and \{0.1--1.0, 1.05--1.5\} (low-$k_{\rm p}$).
}
\label{fig:1}
\end{figure}

\begin{deluxetable*}{lcclcc}[t]
\tabletypesize{\scriptsize}
\tablewidth{0pt}
\tablecaption{List of models with different primordial power spectrum}
\tablehead{
\colhead{PPS model}&
\colhead{$k_{\rm p}$}&
\colhead{$m_{\rm s}$}&
\colhead{$\sigma_8$}&
\colhead{\{$\epsilon_{\rm min}$, $\epsilon_{\rm mean}$, $\epsilon_{\rm max}$\}}&
\colhead{\{$\epsilon_{\rm min}$, $\epsilon_{\rm mean}$, $\epsilon_{\rm max}$\}}\\
\colhead{}&
\colhead{($\hcMpc$)}&
\colhead{}&
\colhead{}&
\colhead{at $z = 9$}&
\colhead{at $z = 7.5$}\\
\colhead{(1)}&
\colhead{(2)}&
\colhead{(3)}&
\colhead{(4)}&
\colhead{(5)}&
\colhead{(6)}
} 
\startdata 
Standard & - & -   & 0.8111    & \{0.17, 0.35, 0.64\} & \{0.25, 0.60, 1.46\} \\
Soft     & 1.0 & 1.5 & 0.8112334 & \{0.10, 0.18, 0.32\} & \{0.17, 0.40, 0.97\} \\
Hard     & 1.0 & 2.0 & 0.8114093 & \{0.07, 0.12, 0.24\} & \{0.13, 0.30, 0.70\} \\
& & & & & \\
Deep     & 5.0 & 1.5 & 0.8111017 & \{0.16, 0.31, 0.65\} & \{0.28, 0.65, 1.49\} \\
Shallow  & 0.3 & 1.2 & 0.8142659 & \{0.09, 0.15, 0.30\} & \{0.15, 0.30, 0.77\} \\
& & & & & \\
Low-$k_{\rm p}$  & {0.1, 0.2, 0.3, 0.5, 1.0} & {1.05, 1.1, 1.2, 1.3, 1.5} & & & \\
\enddata
\tablecomments{
Column (1): model name.
Column (2): pivot scale ($k_{\rm p}$).
Column (3): tilt index ($m_{\rm s}$).
Column (4): root-mean-square matter fluctuation averaged over a sphere of radius $8\,h^{-1}\,{\rm Mpc}$ ($\sigma_8$).
Columns (5) and (6): star formation efficiencies required to exceed the lower limit, center, and upper limit of the observation-limited CCSMD region ($\epsilon_{\rm min}$, $\epsilon_{\rm mean}$, $\epsilon_{\rm max}$) at $z = 9$ and $7.5$.
}
\label{tab:1}
\end{deluxetable*}

\subsection{Primordial power spectrum}

The standard, scale-independent PPS is given by
\begin{equation}
P_{\rm prim}(k) \propto k^{n_{\rm s}}\,,
\label{eq:Pprim_scale-independent}
\end{equation}
whereas the one with enhancement at small scales is given by
\begin{eqnarray}
P_{\rm prim}(k) &\propto& k^{n_{\rm s}} \ ({\rm for}\ k \leq k_{\rm p})\,,\\
&\propto& k_{\rm p}^{n_{\rm s}-m_{\rm s}} \cdot k^{m_{\rm s}} \ ({\rm for}\ k > k_{\rm p})\,.
\label{eq:Pprim_scale-dependent}
\end{eqnarray}
We consider models with different PPS to investigate the dependence of CCSMD and UVLF on the parameters ($k_{\rm p}$ and $m_{\rm s}$):
\begin{itemize}
\item[(1)] We first fix the pivot scale at $k_{\rm p} = 1\,\hcMpc$ ($h$ comoving Mpc$^{-1}$) and adopt two values $m_{\rm s} = 1.5$ and $2.0$, which we call ``soft-tilt'' and ``hard-tilt'', respectively.
We note that our soft-tilt model is in marginally acceptable parameter space from available observational constraints \citep[see Figure~2 in][]{Parashari2023}.
\item[(2)] Next, we examine the influence of the pivot scale by adopting $k_{\rm p} = 0.3$ and $5.0$, which we call shallow model and deep model, respectively.
The latter is motivated by the recent study on the structure of Milky Way satellite galaxies \citep[see Figure~8 in][]{Esteban2023}.
\item[(3)] Finally, we perform a set of simulations for 25 models in total in the ($k_{\rm p}$, $m_{\rm s}$) parameter space with values of $k_{\rm p}/(\hcMpc) = \{0.1$, 0.2, 0.3, 0.5, 1.0\} and $m_{\rm s} = \{1.05$, 1.1, 1.2, 1.3, 1.5\}.
\end{itemize}

For these BTPS models, we adjust the normalization $\sigma_8$ to ensure that the amplitude of the fluctuations at wavelength larger than the pivot scale is the same as in the standard $\Lambda$CDM model (``standard model'').
Figure~\ref{fig:1} shows the resulting matter power spectra at $z = 1089$.

\subsection{Cosmological simulations}

We use the public code \texttt{MUSIC} \citep{Hahn2011} to generate cosmological initial conditions.
We employ $512^3$ dark matter particles in comoving cubes of $L_{\rm box} = 10$, $25$, and $50\,\cMpch$.
Three different simulation volumes are adopted to accurately derive the statistics of nonlinear structure over a wide range of length and mass scales.
The particle mass is $6.52\times10^5$, $1.02\times10^7$, and $8.15\times10^7\,\msun$, respectively.
Dark matter halos with $M_{\rm halo} = 10^8\,\msun$ are resolved by more than $150$ particles in our highest-resolution runs.
Table~\ref{tab:1} summarizes the basic simulation parameters.
We use the parallel $N$-body code \texttt{GADGET-2} \citep{Springel2005} to follow structure formation from redshift $z = 99$ to $0$.

\begin{figure*}[t]
\begin{center}
\includegraphics[width=1.0\linewidth]{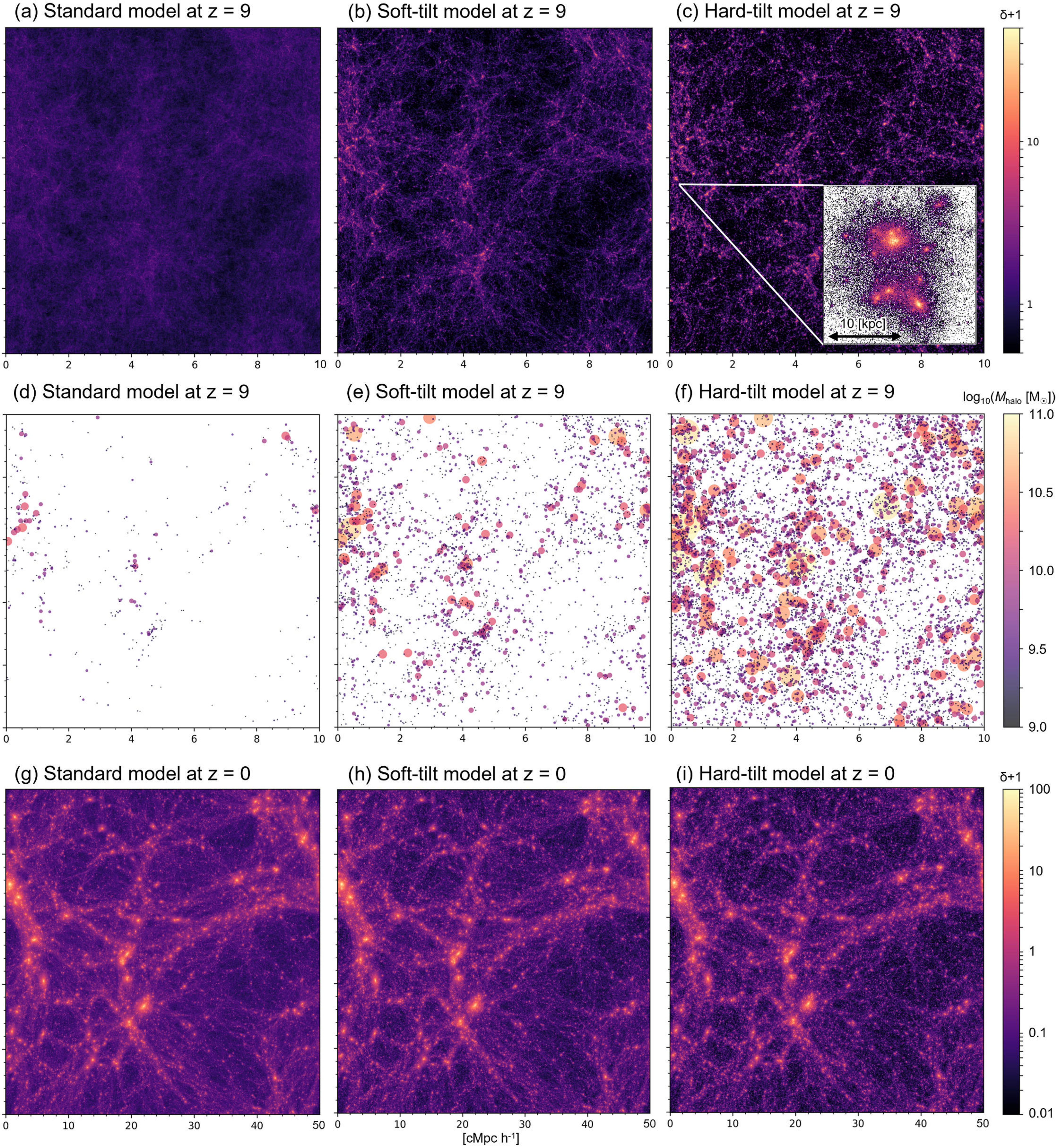}
\end{center}
\caption{
Projected density distributions $\delta+1 = \rho/\bar{\rho}$ and DM halos with $M_{\rm halo} \geq 10^9\,\msun$.
The top and middle panels show the structure at $z = 9$ in a volume of side-length $10\,\cMpch$, whereas the bottom panels show large-scale structure at $z = 0$ with a side-length of $50\,\cMpch$.
The left, center, and right panels are for standard, soft-tilt, and hard-tilt models.
The circle size in the middle panels scales with the halo mass.
The inset in panel (c) is a zoom-in image of one of the most massive halos.
}
\label{fig:2}
\end{figure*}

\begin{figure*}[t]
\begin{center}
\includegraphics[width=1.0\linewidth]{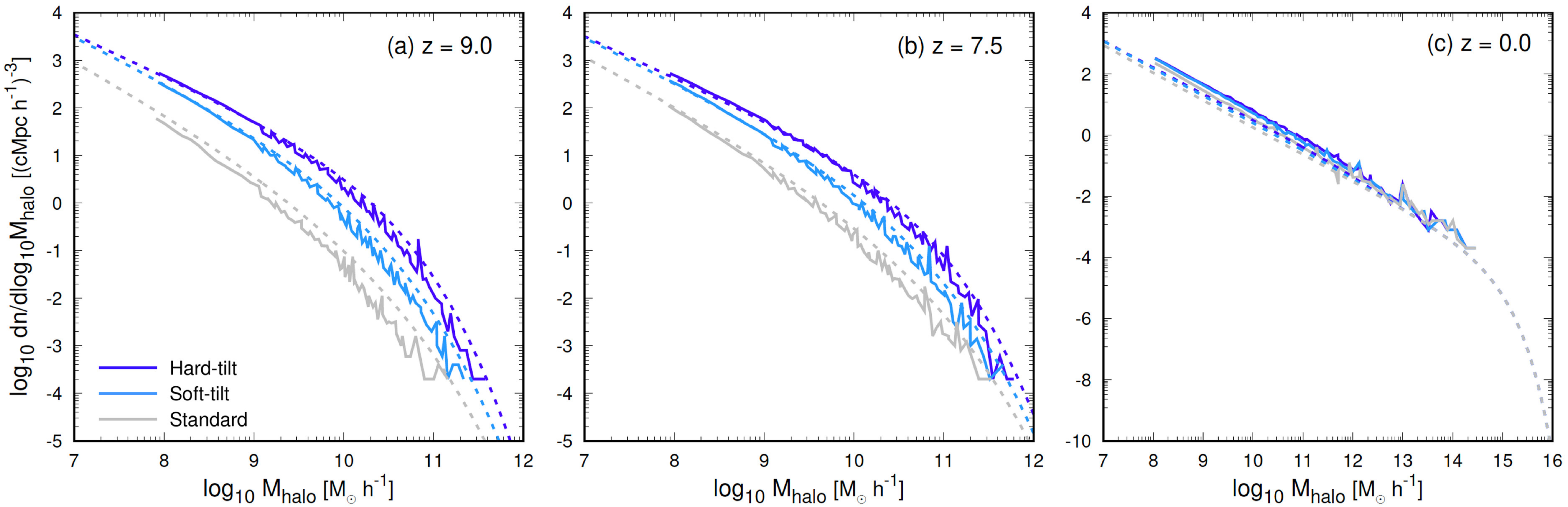}
\end{center}
\caption{
Halo mass functions at $z=9$, $7.5$, and $0$ from left to right.
The solid lines show our simulation results and the dashed lines represent the analytical Sheth–Tormen mass functions \citep{Sheth1999} calculated by the public code \texttt{genmf} \citep{Reed2007}.
The gray, light blue, and dark blue lines are for the standard, soft-tilt, and hard-tilt models.
We construct the simulation data by combining results for different box sizes, $L_{\rm box} = 10$, $25$, and $50\,\cMpch$.
}
\label{fig:3}
\end{figure*}

\subsection{Cumulatice comoving stellar mass density}
\label{sec:method:ccsmd}

We follow \citet{Boylan-Kolchin2023} and \citet{Parashari2023} to compute the cumulative comoving stellar mass density (CCSMD) from the simulation outputs.
We determine the required star formation efficiency ($\epsilon$) to reconcile the recent JWST observation.

We first run a Friends-of-Friends group finder with linking length $b = 0.164 (m/\bar{\rho}(z))^{-3}$, where $m$ is the $N$-body particle mass and $\bar{\rho}(z)$ is the mean density of the universe at redshift $z$, to obtain the halo mass function ${\rm d}n(M, z)/{\rm d}M$.
We combine the halo mass functions from our simulations with different volumes (particle masses) to construct the halo mass function for each model in a wide mass range.
Next, we calculate the cumulative comoving mass density of halos
\begin{equation}
\rho(\geq M_{\rm halo}, z) = \int^{\infty}_{M_{\rm halo}} {\rm d}M M \frac{{\rm d}n(M, z)}{{\rm d}M}\,.
\label{eq:CCSMD}
\end{equation}
Finally, we compute the CCSMD with stellar mass larger than $M_*$ at each redshift, $\rho_*(\geq M_*, z) = \epsilon f_{\rm b} \rho(\geq M_{\rm halo}, z)$. We assume $M_* = \epsilon f_{\rm b} M_{\rm halo}$, where $f_{\rm b} = \Omega_{\rm b} / \Omega_{\rm m} = 0.156$,
and also assume that the star formation efficiency $\epsilon$ ($\leq\!1$) is constant over the redshift range we consider.

\begin{figure*}[t]
\begin{center}
\includegraphics[width=1.0\linewidth]{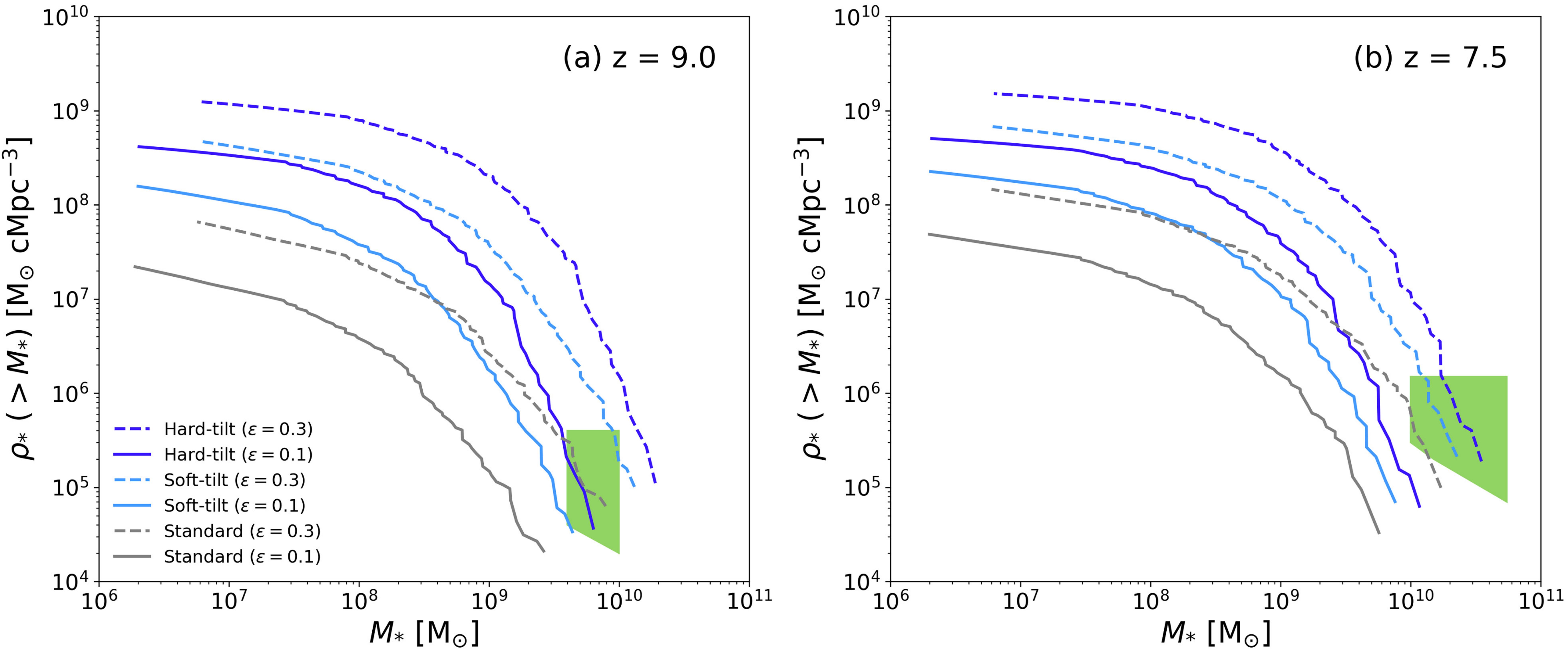}
\end{center}
\caption{
Cumulative comoving stellar mass density (CCSMD) for the standard (gray), soft-tilt (light blue), and hard-tilt (blue) models at $z=9$ (panel a) and $7.5$ (b).
We adopt moderate star formation efficiency of $\epsilon=0.1$ (solid lines) and $0.3$ (dashed).
The green regions are the CCSMD adopted from \cite{Parashari2023} for the observations of \cite{Labbe2023}.
}
\label{fig:4}
\end{figure*}

\section{Stellar Mass Density}
\label{sec:result}

We focus on the cumulative comoving stellar density (CCSMD) as a primary statistic to compare our simulation results and recent JWST observations. 
We first discuss nonlinear structure formation in the BTPS models with different tilt, $m_{\rm s}$.
Then we study the dependence of CCSMD on the other parameter $k_{\rm p}$.
We show that our ``shallow'' models with $k_{\rm p} < 1.0\,\hcMpc$ reproduce the observed CCSMD.
We identify regions in the $k_{\rm p}-m_{\rm s}$ parameter space where the observed CCSMD is reconciled with a reasonable choice of star formation efficiency of $\epsilon = 0.1-0.3$.

\begin{figure*}[t]
\begin{center}
\includegraphics[width=1.0\linewidth]{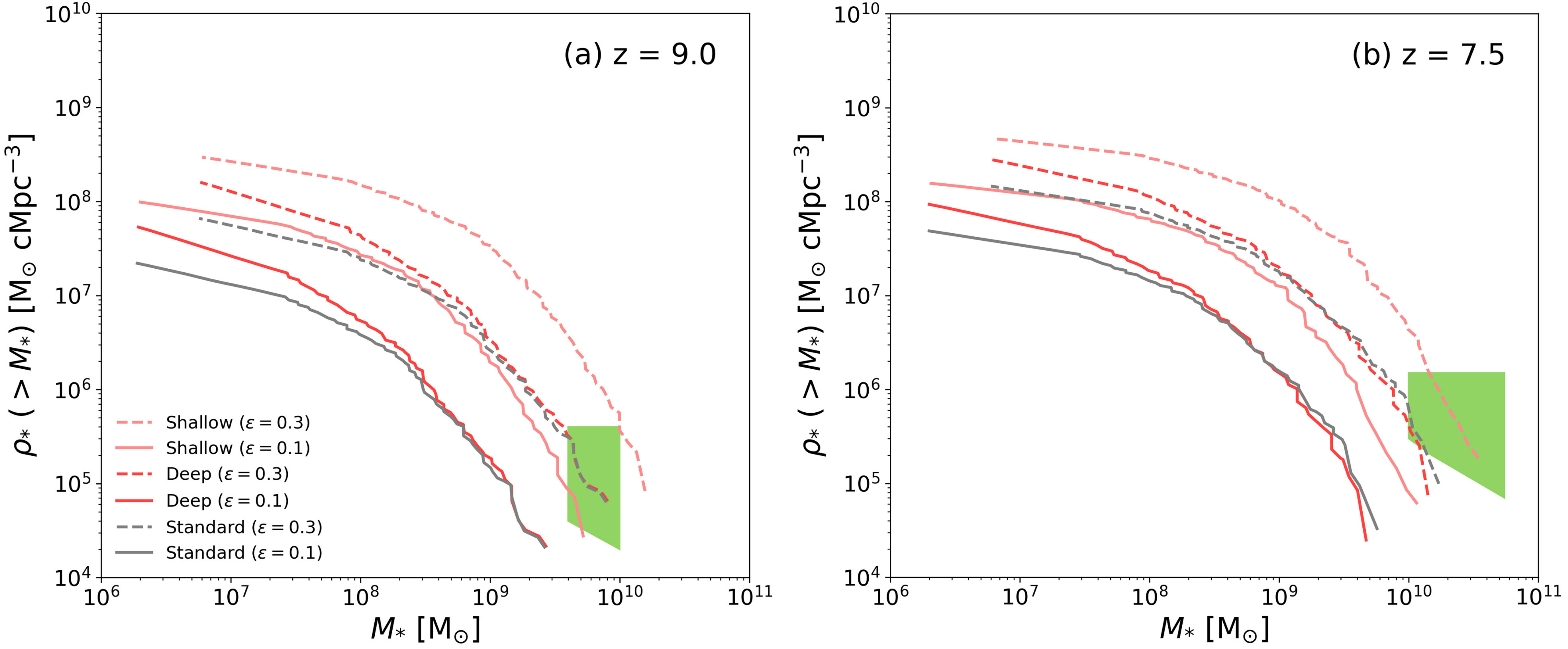}
\end{center}
\caption{
Same as Figure~\ref{fig:4} but for the standard (gray), shallow (light red), and deep (red) models at $z=9$ (panel a) and $7.5$ (b).
}
\label{fig:5}
\end{figure*}

\subsection{Dependence on the tilt, $m_{\rm s}$}
\label{sec:result:tilt}

Figure~\ref{fig:2} shows the projected density distributions for the models with different $m_{\rm s}$.
The enhanced small-scale density fluctuations yield stronger density contrast in the BTPS models than in the standard model; numerous nonlinear objects (halos) are already formed by $z=9$.
We compare the halo mass functions in Figure~\ref{fig:3}, which will be used later to discuss the cumulative stellar mass.
The number density of halos with mass $10^9 < M/\msun < 10^{11}$ differs by an order of magnitude between the hard-tilt model and the standard model.

The middle panels of Figure~\ref{fig:2} show that massive galaxies (halos) are strongly clustered in the BTPS models.
The relative galaxy bias to the underlying dark matter distribution will provide crucial information on the nature of the early galaxy population.
Future wide-field observations of galaxy distribution and clustering by JWST or the Roman Space Telescope will enable us to discriminate theoretical models of galaxy formation \citep{Munoz23}.

The bottom panels of Figure~\ref{fig:2} suggest that the large-scale structure of several tens Mpc in the present-day universe ($z=0$) remains essentially the same.
The enhancement of PPS at the small length scales does not ruin the success of the standard model.
The same holds for the halo mass function at $z=0$ plotted in Figure~\ref{fig:3}(c).

Figure~\ref{fig:4} shows the CCSMDs for the three models at $z=9$ and $7.5$.
There we assume $\epsilon=0.1$ (solid lines) and $0.3$ (dashed lines) with the latter corresponding to the plausible upper limit suggested by recent galaxy formation models \citep{Gribel2017, Tacchella2018, Behroozi2020}.
The green regions indicate the JWST observations adopted from \citet{Parashari2023}, who calculated the CCSMD from the observation of \citet{Labbe2023} with the corresponding spectroscopic updates in two redshift bins $z \in [7, 8.5]$ and $[8.5, 10]$ using the three most massive galaxies.

As has been suggested in the recent literature, a large value of $\epsilon \sim 0.3$ is required for the standard model to reproduce the mean value of the observed CCSMD at $z=9$ (Figure~\ref{fig:4}a).
Even with the ``maximal'' $\epsilon$, the CCSMD does not reach the lower limit of the uncertainty range of the observations at $z=7.5$ (Figure~\ref{fig:4}b).
The CCSMD of the soft-tilt model is about ten and three times larger than the standard model at $z=9$ and $7.5$, respectively.
Thus the soft-tilt model requires a moderate star formation efficiency of $\epsilon \sim 0.1-0.3$ to match the observed CCSMDs at $z=9$ and $7.5$.
Table~\ref{tab:1} summarizes the star formation efficiency required for the CCSMD of each model to exceed the observationally inferred lower limit ($\epsilon_{\rm min}$), mean ($\epsilon_{\rm mean}$), and upper limit ($\epsilon_{\rm max}$).

\subsection{Dependence on the pivot scale, $k_{\rm p}$}
\label{sec:result:pivot}

We have set the fiducial pivot scale at $k_{\rm p} = 1$ somewhat arbitrarily.
To examine the effect of the pivot scale on the CCSMD, we consider ``shallow'' and ``deep'' models with $k_{\rm p}=0.3$ and $5$, respectively.
Figure~\ref{fig:5} shows that the resulting CCSMDs are nearly the same as that of the standard model at the massive end, and the values are slightly larger at small masses.
Thus the deep model requires a very high star formation efficiency to reconcile with the JWST observations, similar to the standard model.

Interestingly, our shallow model successfully increases the CCSMD to match the observation, thus relaxing the star formation efficiency requirement.
Even though the enhancement on the small-scale power appears small compared to the other models (soft, hard, deep), the shallow model effectively increases the abundance of galactic dark halos at the relevant mass and redshift.
Note that a small bump in the power spectrum around $k \sim 1$ could also explain the apparently rapid galaxy formation in the early universe \citep{Padmanabhan&Loeb2023}.

\subsection{Parameter survey for low-$k_{\rm p}$}
\label{sec:result:survey}

\begin{figure}[t]
\begin{center}
\includegraphics[width=1.0\linewidth]{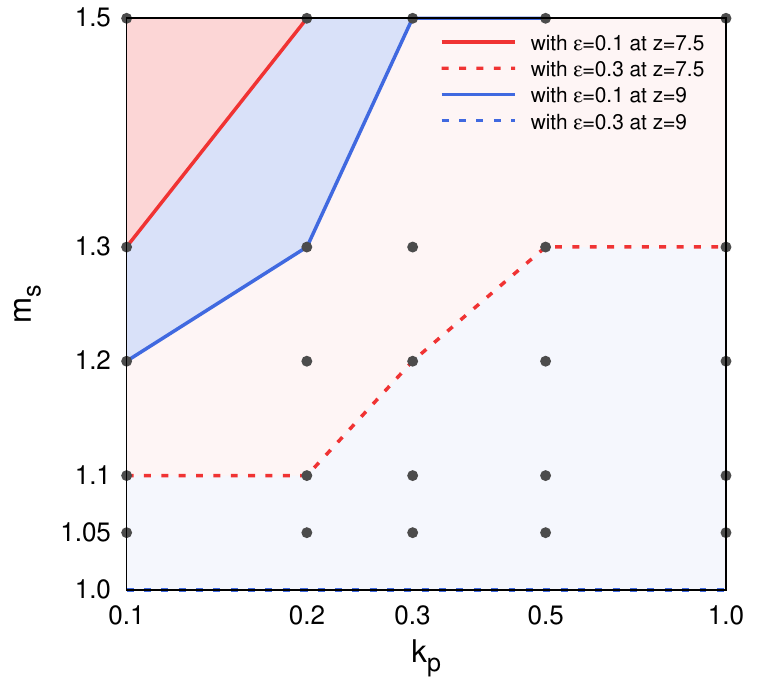}
\end{center}
\caption{
Summary of our simulation-based constraints in $k_{\rm p}$-$m_{\rm s}$ space.
We calculate the CCSMDs for each model at $z=9$ (blue) and $7.5$ (red) with $\epsilon = 0.1$ (solid) and $0.3$ (dashed).
We divide the ``accepted'' and constrained regions by the dashed and solid lines.
Above the line for each model, the CCSMD matches the mean of the observed range (green regions in Figure~\ref{fig:4}).
}
\label{fig:6}
\end{figure}

Motivated by the success of our shallow model, we explore a wider range of models with low $k_{\rm p}$ and derive an observationally allowed parameter range of the power spectrum.
Figure~\ref{fig:6} summarizes the results on the $k_{\rm p}$-$m_{\rm s}$ plane.
We accept a model if the CCSMD value exceeds the central value of the JWST observations (shaded regions in Figures~\ref{fig:4} and \ref{fig:5}).
We judge the models at two different redshifts ($z=7.5$ and $9$) and with two different star formation efficiencies ($\epsilon=0.1$ and $0.3$) independently.
In Figure~\ref{fig:6}, models on and above the respective boundary lines are ``acceptable''.
For instance, models with $m_{\rm s} = 1.1$ and with $k_{\rm p} < 0.2$ successfully reproduce the observed CCSMD and hence ``accepted''.

We argue that a slight enhancement of the PPS in the previously unexplored range of $k_{\rm p}=0.1-0.2$ could explain the rapid formation of galaxies in the early Universe.
Among low-$k_{\rm p}$ BTPS models, we select one model with $k_{\rm p} = 0.2\,\hcMpc$ and $m_{\rm s}=1.3$ as the ``minimal shallow'' model that could explain available observations with the modest change of the PPS.

\section{Luminosity Function}
\label{sec:UVLF}

The ultra-violet luminosity function (UVLF) may be another useful probe of the small-scale power, although the resulting constraints depend much on the adopted models of galaxy formation and mass-to-light conversion \citep{Munoz23}.

We follow \citet{Sabti2022PRD, Sabti2022ApJ} to compute the UVLF from the simulation outputs.
To simplify the model comparison, we adopt the best-fit parameter values in their fiducial model obtained from the Hubble Space Telescope data.
We first calculate the halo mass accretion rate at redshift $z$ by comparing halo masses of the same halo identified by the Friends-of-Friends algorithm at different redshift with $dz = 0.1$ as $\dot{M}_{\rm h}(z) = (M_{\rm h}(z)-M_{\rm h}(z+dz))/dz$. 
Next, we translate the halo mass accretion rate to the star formation rate (SFR) by assuming a double power-law relation,
\begin{equation}
f_*(z, M_{\rm h}) = \frac{\dot{M}_{\rm *}}{\dot{M}_{\rm h}} = \frac{\epsilon_*}{\left(\frac{M_{\rm h}}{M_{\rm c}}\right)^{\alpha_*} + \left(\frac{M_{\rm h}}{M_{\rm c}}\right)^{\beta_*}}\,,
\label{eq:SFR}
\end{equation}
with the following forms,
\begin{eqnarray}
\alpha_*(z) &=& \alpha_* \label{eq:SFRform-1} \\
\beta_*(z) &=& \beta_* \label{eq:SFRform-2} \\
\log_{10} \epsilon_*(z) &=& \epsilon_*^{\rm s} \log_{10} \left(\frac{1+z}{1+6}\right) + \epsilon_*^{\rm i} \label{eq:SFRform-3} \\ 
\log_{10} M_{\rm c}(z)/\msun &=& M_{\rm c}^{\rm s} \log_{10} \left(\frac{1+z}{1+6}\right) + M_{\rm c}^{\rm i}\,. \label{eq:SFRform-4}
\end{eqnarray}
The best-fit values of six free parameters are \{$\alpha_*$ $\beta_*$, $\epsilon_*^{\rm s}$, $\epsilon_*^{\rm i}$ $M_{\rm c}^{\rm s}$, $M_{\rm c}^{\rm i}$\} = \{-0.594, 0.611, -1.96, -2.17, 2.95, 12.1\}.

We relate the UV luminosity of a galaxy to the SFR as \citep{Kennicutt1998, Madau1998},
\begin{equation}
\dot{M}_{\rm *} = \kappa_{\rm UV} L_{\rm UV}\,,
\label{eq:SFRtoLuv}
\end{equation}
with a popular choice of $\kappa_{\rm UV} = 1.15 \times 10^{-28}\,\msun\,{\rm s}\,{\rm erg}^{-1}\,{\rm yr}^{-1}$ \citep{MadauDickinson2014}.
We then convert the UV luminosity to the absolute magnitude \citep{OkeGunn1983},
\citep{Kennicutt1998, Madau1998},
\begin{equation}
\log_{10}\left(\frac{L_{\rm UV}}{{\rm erg\,s^{-1}}}\right) = 0.4 (51.63 - M_{\rm UV})\,.
\label{eq:LuvtoMuv}
\end{equation}

Using the above formulae, we calculate the absolute UV magnitudes ($M_{\rm UV}$) from the halo mass accretion rates ($\dot{M}_{\rm h}$) obtained from the simulations.
We calculate the UVLF for each model at different redshifts.

\begin{figure}[t]
\begin{center}
\includegraphics[width=1.0\linewidth]{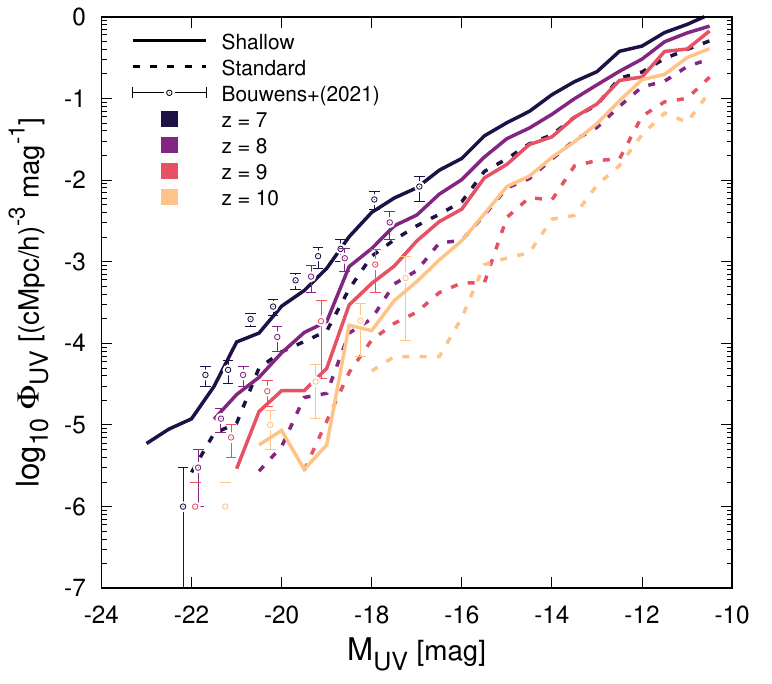}
\end{center}
\caption{
The galaxy UVLF for our shallow model (solid lines) and for the standard model (dashed) at $z = 7, 8, 9$, and $10$.
We also plot the observational data of \citet{Bouwens2021}.
}
\label{fig:7}
\end{figure}

Figure~\ref{fig:7} compares the recent observation \citep{Bouwens2021} to the calculated UVLFs for the minimal shallow ($k_{\rm p} = 0.2\,\hcMpc$ and $m_{\rm s} = 1.3$) and standard models.
Compared to the standard model, the minimal shallow model increases the UVLF, but
does not show significant deviations from the observed range.
One reason is that the BTPS model promotes formation of many low-mass galaxies but does not affect the formation of massive galaxies, which contribute to the UVLF in the relevant luminosity range.
Interestingly, the minimal shallow model is closer to the observational data than the standard one with the model employed here.
To the extent of this comparison, the current UVLF observations do not constrain the shallow tilt BTPS model.

We note that the star formation efficiencies adopted in the above UVLF model and our CCSMD model may not necessarily be consistent. 
The CCSMDs are calculated by assuming a fraction $\epsilon$ of the gas mass is converted into stars, whereas the UVLF model assumes a fraction $f_*$ of the accreted {\it halo} mass is converted to stars at a given instant.
It is not straightforward to compare the two efficiencies directly, for one is related to the resulting stellar mass by a given epoch and the other is the instantaneous formation rate.
Also, the UVLF model we consider here involves an effective mass-to-light ratio (Equation~\ref{eq:SFRtoLuv}) which is uncertain and depends critically on the assumed stellar initial mass function.
It is beyond the scope of the present paper to develop a consistent galaxy formation model, but it would be interesting to explore, for example, semi-analytic models of galaxy formation that treat the star formation process in a self-consistent manner.
We also note the effective halo mass range probed by the JWST CCSMD is different from that of the HST UVLF function, the latter of which probes overall larger halos with $M_{\rm halo} = 10^{11} - 10^{12} M_{\odot}$ and hence the effective power at lower wavenumber \citep{Munoz23}.

\section{Discussion}
\label{sec:conc}

The result of our cosmological simulations is largely consistent with the estimate of \cite{Parashari2023} based on an analytic halo mass function. 
Our simulations confirm that the standard $\Lambda$CDM model requires unrealistically high star formation efficiencies of $\epsilon > 0.3$ to reconcile the observed CCSMD. 
Although there have already been proposals for galaxy formation physics that can realize a high star formation efficiency \citep[e.g.,][]{Dekel23}, a slight modification of the PPS may be another promising solution that alleviates the need for significant deviation from the currently popular galaxy formation models.

Interestingly, recent JWST observations also show that there are many galaxies with clumpy structures and galaxies in the process of mergers in proto-cluster environments at $z=7-9$ \citep{Hashimoto23, Hainline23}. 
The nonlinear structure forms early in the BTPS model and assembles rapidly via mergers.
Thus, galaxies at $z=7-9$ tend to appear clumpy, as seen in Figure~\ref{fig:2}.

The PPS strongly affects the formation epoch and the properties of the Population III stars. 
\citet{Hirano2015blue} run a set of hydrodynamics simulations starting from nearly the same BTPS and find that very massive stars with mass exceeding $100\,\msun$ are formed early.
Recently, \citet{Xing2023} discovered a metal-poor star that shows definite chemical signatures of the so-called pair-instability supernova caused by a very massive star early in the formation history of the Milky Way.
Formation efficiency and epoch of very massive stars may provide another hint on the small-scale density fluctuations.

The existence of super-massive black holes in the early universe generally suggests rapid growth of structure and formation of appropriate seed black holes at early epochs \cite[e.g.,][]{Inayoshi20}.
\citet{Larson23} discovered a massive black hole candidate at $z=8.6$.
Stellar-mass black holes are formed as remnants of massive Population III stars as early as $z \sim 50-100$ in the BTPS model, leaving enough time for the seeds to grow by mass accretion to massive black holes by $z=8$.
Frequent halo mergers realized in the BTPS model may also enhance the formation of massive black holes in the early universe \citep{Wise19, Regan23}.

Tight constraints can be placed on the slope or the amplitude of PPS at sub-galactic length scales from observations of the history of cosmic reionization.
Contribution to early reionization from individual Population III stars is severely limited by the measurement of the Thomson optical depth by Planck \citep{Visbal15}.
\citet{Minoda23} derive constraints on the ``running'' of the PPS from CMB and recent observations of high-redshift galaxies, which still allow both negative (red) and positive (blue) running to a modest degree.
Enhanced small-scale power promotes early star formation and also increases inter-galactic medium clumping (see Figure~\ref{fig:2}).
The photon escape fraction of galaxies, an important quantity in the study of cosmic reionization, is uncertain and likely affected by the gas clumping and the abundance of substructure within early galaxies \citep{Yeh23}. 
Also, radiative and mechanical feedback effects from Population III stars may actually suppress, at least temporarily, star formation in small galaxies that form later around $z\sim10-15$ \citep{Bromm11}. 
Overall, there still remains substantial uncertainty in both observations of reionization history \citep{WMAP9, PlanckCollaboration2020, Forconi23} and in astrophysical modelling \citep{Fialkov22}.
It would be necessary to perform detailed numerical simulations of galaxy formation and reionization to place constraints on the BTPS model.

Structure of galactic halos may provide another clue on the small-scale power.
Since low-mass satellite galaxies are too faint to be observed in the distant Universe, observations of the nearby Universe are often used to infer the small-scale density fluctuations  \citep[e.g.,][]{Esteban2023}.
Interestingly, \cite{Honma2023} recently reported that the abundance of ultrafaint dwarf galaxies in the Milky Way is {\it higher} than the theoretical prediction from the standard $\Lambda$CDM model with a single power-law PPS, posing a ``{\it too many satellites problem}''. 
Detailed galaxy formation models combined with a broad class of observations of sub-galactic objects are necessary to understand the formation and evolution of small-scale structures and to place tight constraints on the PPS.
\citet{Tkachev2024} consider a modified matter power spectrum with an approximately Gaussian bump at sub-Mpc scales and study the effect on the halo mass function.
They argue that observations of high-redshift galaxies are crucial to infer the feature of the small-scale power spectrum.

JWST has opened a new window into the distant universe.
Future observations by JWST and other wide-field cosmology surveys will reveal the physical properties and the formation history of galaxy {\it populations} at high redshift.
Upcoming data may ultimately provide new insight into the early universe physics that generates the primordial density fluctuations from which the first galaxies are formed.

\begin{acknowledgments}
Numerical computations were performed on Cray XC50 at CfCA in National Astronomical Observatory of Japan and Yukawa-21 at YITP in Kyoto University.
Numerical analyses were in part carried out on the analysis servers at CfCA in National Astronomical Observatory of Japan.
This work was supported by JSPS KAKENHI Grant Numbers JP21K13960, JP21H01123, and JP22H01259 (S.H.), and MEXT as ``Program for Promoting Researches on the Supercomputer Fugaku'' (Structure and Evolution of the Universe Unraveled by Fusion of Simulation and AI; Grant Number JPMXP1020230406, Project ID hp230204) (S.H. and N.Y).
\end{acknowledgments}

\bibliography{ms}{}

\begin{thebibliography}{}
\expandafter\ifx\csname natexlab\endcsname\relax\def\natexlab#1{#1}\fi
\providecommand{\url}[1]{\href{#1}{#1}}
\providecommand{\dodoi}[1]{doi:~\href{http://doi.org/#1}{\nolinkurl{#1}}}
\providecommand{\doeprint}[1]{\href{http://ascl.net/#1}{\nolinkurl{http://ascl.net/#1}}}
\providecommand{\doarXiv}[1]{\href{https://arxiv.org/abs/#1}{\nolinkurl{https://arxiv.org/abs/#1}}}

\bibitem[{{Behroozi} {et~al.}(2020){Behroozi}, {Conroy}, {Wechsler}, {Hearin}, {Williams}, {Moster}, {Yung}, {Somerville}, {Gottl{\"o}ber}, {Yepes}, \& {Endsley}}]{Behroozi2020}
{Behroozi}, P., {Conroy}, C., {Wechsler}, R.~H., {et~al.} 2020, \mnras, 499, 5702, \dodoi{10.1093/mnras/staa3164}

\bibitem[{{Bouwens} {et~al.}(2021){Bouwens}, {Oesch}, {Stefanon}, {Illingworth}, {Labb{\'e}}, {Reddy}, {Atek}, {Montes}, {Naidu}, {Nanayakkara}, {Nelson}, \& {Wilkins}}]{Bouwens2021}
{Bouwens}, R.~J., {Oesch}, P.~A., {Stefanon}, M., {et~al.} 2021, \aj, 162, 47, \dodoi{10.3847/1538-3881/abf83e}

\bibitem[{{Boylan-Kolchin}(2023)}]{Boylan-Kolchin2023}
{Boylan-Kolchin}, M. 2023, Nature Astronomy, \dodoi{10.1038/s41550-023-01937-7}

\bibitem[{{Bromm} \& {Yoshida}(2011)}]{Bromm11}
{Bromm}, V., \& {Yoshida}, N. 2011, \araa, 49, 373, \dodoi{10.1146/annurev-astro-081710-102608}

\bibitem[{{Bullock} \& {Boylan-Kolchin}(2017)}]{Bullock2017}
{Bullock}, J.~S., \& {Boylan-Kolchin}, M. 2017, \araa, 55, 343, \dodoi{10.1146/annurev-astro-091916-055313}

\bibitem[{{Clesse} \& {Garc{\'\i}a-Bellido}(2015)}]{Clesse2015}
{Clesse}, S., \& {Garc{\'\i}a-Bellido}, J. 2015, \prd, 92, 023524, \dodoi{10.1103/PhysRevD.92.023524}

\bibitem[{{Covi} \& {Lyth}(1999)}]{Covy1999}
{Covi}, L., \& {Lyth}, D.~H. 1999, \prd, 59, 063515, \dodoi{10.1103/PhysRevD.59.063515}

\bibitem[{{Dekel} {et~al.}(2023){Dekel}, {Sarkar}, {Birnboim}, {Mandelker}, \& {Li}}]{Dekel23}
{Dekel}, A., {Sarkar}, K.~C., {Birnboim}, Y., {Mandelker}, N., \& {Li}, Z. 2023, \mnras, 523, 3201, \dodoi{10.1093/mnras/stad1557}

\bibitem[{{Esteban} {et~al.}(2023){Esteban}, {Peter}, \& {Kim}}]{Esteban2023}
{Esteban}, I., {Peter}, A. H.~G., \& {Kim}, S.~Y. 2023, arXiv e-prints, arXiv:2306.04674, \dodoi{10.48550/arXiv.2306.04674}

\bibitem[{{Fialkov}(2022)}]{Fialkov22}
{Fialkov}, A. 2022, in The Fifteenth Marcel Grossmann Meeting on General Relativity. Edited by E. S. Battistelli, ed. E.~S. {Battistelli}, R.~T. {Jantzen}, \& R.~{Ruffini}, 1067--1073, \dodoi{10.1142/9789811258251_0149}

\bibitem[{{Finkelstein} {et~al.}(2022){Finkelstein}, {Bagley}, {Haro}, {Dickinson}, {Ferguson}, {Kartaltepe}, {Papovich}, {Burgarella}, {Kocevski}, {Huertas-Company}, {Iyer}, {Koekemoer}, {Larson}, {P{\'e}rez-Gonz{\'a}lez}, {Rose}, {Tacchella}, {Wilkins}, {Chworowsky}, {Medrano}, {Morales}, {Somerville}, {Yung}, {Fontana}, {Giavalisco}, {Grazian}, {Grogin}, {Kewley}, {Kirkpatrick}, {Kurczynski}, {Lotz}, {Pentericci}, {Pirzkal}, {Ravindranath}, {Ryan}, {Trump}, {Yang}, {Almaini}, {Amor{\'\i}n}, {Annunziatella}, {Backhaus}, {Barro}, {Behroozi}, {Bell}, {Bhatawdekar}, {Bisigello}, {Bromm}, {Buat}, {Buitrago}, {Calabr{\`o}}, {Casey}, {Castellano}, {Ch{\'a}vez Ortiz}, {Ciesla}, {Cleri}, {Cohen}, {Cole}, {Cooke}, {Cooper}, {Cooray}, {Costantin}, {Cox}, {Croton}, {Daddi}, {Dav{\'e}}, {de La Vega}, {Dekel}, {Elbaz}, {Estrada-Carpenter}, {Faber}, {Fern{\'a}ndez}, {Finkelstein}, {Freundlich}, {Fujimoto}, {Garc{\'\i}a-Argum{\'a}nez}, {Gardner}, {Gawiser}, {G{\'o}mez-Guijarro}, {Guo}, {Hamblin}, {Hamilton}, {Hathi},
  {Holwerda}, {Hirschmann}, {Hutchison}, {Jaskot}, {Jha}, {Jogee}, {Juneau}, {Jung}, {Kassin}, {Bail}, {Leung}, {Lucas}, {Magnelli}, {Mantha}, {Matharu}, {McGrath}, {McIntosh}, {Merlin}, {Mobasher}, {Newman}, {Nicholls}, {Pandya}, {Rafelski}, {Ronayne}, {Santini}, {Seill{\'e}}, {Shah}, {Shen}, {Simons}, {Snyder}, {Stanway}, {Straughn}, {Teplitz}, {Vanderhoof}, {Vega-Ferrero}, {Wang}, {Weiner}, {Willmer}, {Wuyts}, {Zavala}, \& {Ceers Team}}]{Finkelstein2022}
{Finkelstein}, S.~L., {Bagley}, M.~B., {Haro}, P.~A., {et~al.} 2022, \apjl, 940, L55, \dodoi{10.3847/2041-8213/ac966e}

\bibitem[{{Forconi} {et~al.}(2023){Forconi}, {Ruchika}, {Melchiorri}, {Mena}, \& {Menci}}]{Forconi23}
{Forconi}, M., {Ruchika}, {Melchiorri}, A., {Mena}, O., \& {Menci}, N. 2023, \jcap, 2023, 012, \dodoi{10.1088/1475-7516/2023/10/012}

\bibitem[{{Germani} \& {Prokopec}(2017)}]{Germani2017}
{Germani}, C., \& {Prokopec}, T. 2017, Physics of the Dark Universe, 18, 6, \dodoi{10.1016/j.dark.2017.09.001}

\bibitem[{{Gong} \& {Sasaki}(2011)}]{Gong2011}
{Gong}, J.-O., \& {Sasaki}, M. 2011, \jcap, 2011, 028, \dodoi{10.1088/1475-7516/2011/03/028}

\bibitem[{{Gribel} {et~al.}(2017){Gribel}, {Miranda}, \& {Williams Vilas-Boas}}]{Gribel2017}
{Gribel}, C., {Miranda}, O.~D., \& {Williams Vilas-Boas}, J. 2017, \apj, 849, 108, \dodoi{10.3847/1538-4357/aa921a}

\bibitem[{{Hahn} \& {Abel}(2011)}]{Hahn2011}
{Hahn}, O., \& {Abel}, T. 2011, \mnras, 415, 2101, \dodoi{10.1111/j.1365-2966.2011.18820.x}

\bibitem[{{Hainline} {et~al.}(2023){Hainline}, {Johnson}, {Robertson}, {Tacchella}, {Helton}, {Sun}, {Eisenstein}, {Simmonds}, {Topping}, {Whitler}, {Willmer}, {Rieke}, {Suess}, {Hviding}, {Cameron}, {Alberts}, {Baker}, {Bhatawdekar}, {Boyett}, {Bunker}, {Carniani}, {Charlot}, {Chen}, {Curti}, {Curtis-Lake}, {D'Eugenio}, {Egami}, {Endsley}, {Hausen}, {Ji}, {Looser}, {Lyu}, {Maiolino}, {Nelson}, {Puskas}, {Rawle}, {Sandles}, {Saxena}, {Smit}, {Stark}, {Williams}, {Willott}, \& {Witstok}}]{Hainline23}
{Hainline}, K.~N., {Johnson}, B.~D., {Robertson}, B., {et~al.} 2023, arXiv e-prints, arXiv:2306.02468, \dodoi{10.48550/arXiv.2306.02468}

\bibitem[{{Hashimoto} {et~al.}(2023){Hashimoto}, {{\'A}lvarez-M{\'a}rquez}, {Fudamoto}, {Colina}, {Inoue}, {Nakazato}, {Ceverino}, {Yoshida}, {Costantin}, {Sugahara}, {G{\'o}mez}, {Blanco-Prieto}, {Mawatari}, {Arribas}, {Marques-Chaves}, {Pereira-Santaella}, {Bakx}, {Hagimoto}, {Hashigaya}, {Matsuo}, {Tamura}, {Usui}, \& {Ren}}]{Hashimoto23}
{Hashimoto}, T., {{\'A}lvarez-M{\'a}rquez}, J., {Fudamoto}, Y., {et~al.} 2023, \apjl, 955, L2, \dodoi{10.3847/2041-8213/acf57c}

\bibitem[{{Hinshaw} {et~al.}(2013){Hinshaw}, {Larson}, {Komatsu}, {Spergel}, {Bennett}, {Dunkley}, {Nolta}, {Halpern}, {Hill}, {Odegard}, {Page}, {Smith}, {Weiland}, {Gold}, {Jarosik}, {Kogut}, {Limon}, {Meyer}, {Tucker}, {Wollack}, \& {Wright}}]{WMAP9}
{Hinshaw}, G., {Larson}, D., {Komatsu}, E., {et~al.} 2013, \apjs, 208, 19, \dodoi{10.1088/0067-0049/208/2/19}

\bibitem[{{Hirano} {et~al.}(2015){Hirano}, {Zhu}, {Yoshida}, {Spergel}, \& {Yorke}}]{Hirano2015blue}
{Hirano}, S., {Zhu}, N., {Yoshida}, N., {Spergel}, D., \& {Yorke}, H.~W. 2015, \apj, 814, 18, \dodoi{10.1088/0004-637X/814/1/18}

\bibitem[{{Hlozek} {et~al.}(2012){Hlozek}, {Dunkley}, {Addison}, {Appel}, {Bond}, {Sofia Carvalho}, {Das}, {Devlin}, {D{\"u}nner}, {Essinger-Hileman}, {Fowler}, {Gallardo}, {Hajian}, {Halpern}, {Hasselfield}, {Hilton}, {Hincks}, {Hughes}, {Irwin}, {Klein}, {Kosowsky}, {Marriage}, {Marsden}, {Menanteau}, {Moodley}, {Niemack}, {Nolta}, {Page}, {Parker}, {Partridge}, {Rojas}, {Sehgal}, {Sherwin}, {Sievers}, {Spergel}, {Staggs}, {Swetz}, {Switzer}, {Thornton}, \& {Wollack}}]{Hlozek2012}
{Hlozek}, R., {Dunkley}, J., {Addison}, G., {et~al.} 2012, \apj, 749, 90, \dodoi{10.1088/0004-637X/749/1/90}

\bibitem[{{Homma} {et~al.}(2023){Homma}, {Chiba}, {Komiyama}, {Tanaka}, {Okamoto}, {Tanaka}, {Ishigaki}, {Hayashi}, {Arimoto}, {Lupton}, {Strauss}, {Miyazaki}, {Wang}, \& {Murayama}}]{Honma2023}
{Homma}, D., {Chiba}, M., {Komiyama}, Y., {et~al.} 2023, arXiv e-prints, arXiv:2311.05439, \dodoi{10.48550/arXiv.2311.05439}

\bibitem[{{Inayoshi} {et~al.}(2020){Inayoshi}, {Visbal}, \& {Haiman}}]{Inayoshi20}
{Inayoshi}, K., {Visbal}, E., \& {Haiman}, Z. 2020, \araa, 58, 27, \dodoi{10.1146/annurev-astro-120419-014455}

\bibitem[{{Inman} \& {Kohri}(2023)}]{Inman2023}
{Inman}, D., \& {Kohri}, K. 2023, \prd, 107, 123513, \dodoi{10.1103/PhysRevD.107.123513}

\bibitem[{{Kennicutt}(1998)}]{Kennicutt1998}
{Kennicutt}, Robert~C., J. 1998, \araa, 36, 189, \dodoi{10.1146/annurev.astro.36.1.189}

\bibitem[{{Labb{\'e}} {et~al.}(2023){Labb{\'e}}, {van Dokkum}, {Nelson}, {Bezanson}, {Suess}, {Leja}, {Brammer}, {Whitaker}, {Mathews}, {Stefanon}, \& {Wang}}]{Labbe2023}
{Labb{\'e}}, I., {van Dokkum}, P., {Nelson}, E., {et~al.} 2023, \nat, 616, 266, \dodoi{10.1038/s41586-023-05786-2}

\bibitem[{{Larson} {et~al.}(2023){Larson}, {Finkelstein}, {Kocevski}, {Hutchison}, {Trump}, {Arrabal Haro}, {Bromm}, {Cleri}, {Dickinson}, {Fujimoto}, {Kartaltepe}, {Koekemoer}, {Papovich}, {Pirzkal}, {Tacchella}, {Zavala}, {Bagley}, {Behroozi}, {Champagne}, {Cole}, {Jung}, {Morales}, {Yang}, {Zhang}, {Zitrin}, {Amor{\'\i}n}, {Burgarella}, {Casey}, {Ch{\'a}vez Ortiz}, {Cox}, {Chworowsky}, {Fontana}, {Gawiser}, {Grazian}, {Grogin}, {Harish}, {Hathi}, {Hirschmann}, {Holwerda}, {Juneau}, {Leung}, {Lucas}, {McGrath}, {P{\'e}rez-Gonz{\'a}lez}, {Rigby}, {Seill{\'e}}, {Simons}, {de La Vega}, {Weiner}, {Wilkins}, {Yung}, \& {Ceers Team}}]{Larson23}
{Larson}, R.~L., {Finkelstein}, S.~L., {Kocevski}, D.~D., {et~al.} 2023, \apjl, 953, L29, \dodoi{10.3847/2041-8213/ace619}

\bibitem[{{Madau} \& {Dickinson}(2014)}]{MadauDickinson2014}
{Madau}, P., \& {Dickinson}, M. 2014, \araa, 52, 415, \dodoi{10.1146/annurev-astro-081811-125615}

\bibitem[{{Madau} {et~al.}(1998){Madau}, {Pozzetti}, \& {Dickinson}}]{Madau1998}
{Madau}, P., {Pozzetti}, L., \& {Dickinson}, M. 1998, \apj, 498, 106, \dodoi{10.1086/305523}

\bibitem[{{Martin} \& {Brandenberger}(2001)}]{Martin2001}
{Martin}, J., \& {Brandenberger}, R.~H. 2001, \prd, 63, 123501, \dodoi{10.1103/PhysRevD.63.123501}

\bibitem[{{Minoda} {et~al.}(2023){Minoda}, {Yoshiura}, \& {Takahashi}}]{Minoda23}
{Minoda}, T., {Yoshiura}, S., \& {Takahashi}, T. 2023, \prd, 108, 123542, \dodoi{10.1103/PhysRevD.108.123542}

\bibitem[{{Mu{\~n}oz} {et~al.}(2023){Mu{\~n}oz}, {Mirocha}, {Furlanetto}, \& {Sabti}}]{Munoz23}
{Mu{\~n}oz}, J.~B., {Mirocha}, J., {Furlanetto}, S., \& {Sabti}, N. 2023, \mnras, 526, L47, \dodoi{10.1093/mnrasl/slad115}

\bibitem[{{Oke} \& {Gunn}(1983)}]{OkeGunn1983}
{Oke}, J.~B., \& {Gunn}, J.~E. 1983, \apj, 266, 713, \dodoi{10.1086/160817}

\bibitem[{{Padmanabhan} \& {Loeb}(2023)}]{Padmanabhan&Loeb2023}
{Padmanabhan}, H., \& {Loeb}, A. 2023, \apjl, 953, L4, \dodoi{10.3847/2041-8213/acea7a}

\bibitem[{{Parashari} \& {Laha}(2023)}]{Parashari2023}
{Parashari}, P., \& {Laha}, R. 2023, \mnras, 526, L63, \dodoi{10.1093/mnrasl/slad107}

\bibitem[{{Planck Collaboration} {et~al.}(2020){Planck Collaboration}, {Aghanim}, {Akrami}, {Ashdown}, {Aumont}, {Baccigalupi}, {Ballardini}, {Banday}, {Barreiro}, {Bartolo}, {Basak}, {Battye}, {Benabed}, {Bernard}, {Bersanelli}, {Bielewicz}, {Bock}, {Bond}, {Borrill}, {Bouchet}, {Boulanger}, {Bucher}, {Burigana}, {Butler}, {Calabrese}, {Cardoso}, {Carron}, {Challinor}, {Chiang}, {Chluba}, {Colombo}, {Combet}, {Contreras}, {Crill}, {Cuttaia}, {de Bernardis}, {de Zotti}, {Delabrouille}, {Delouis}, {Di Valentino}, {Diego}, {Dor{\'e}}, {Douspis}, {Ducout}, {Dupac}, {Dusini}, {Efstathiou}, {Elsner}, {En{\ss}lin}, {Eriksen}, {Fantaye}, {Farhang}, {Fergusson}, {Fernandez-Cobos}, {Finelli}, {Forastieri}, {Frailis}, {Fraisse}, {Franceschi}, {Frolov}, {Galeotta}, {Galli}, {Ganga}, {G{\'e}nova-Santos}, {Gerbino}, {Ghosh}, {Gonz{\'a}lez-Nuevo}, {G{\'o}rski}, {Gratton}, {Gruppuso}, {Gudmundsson}, {Hamann}, {Handley}, {Hansen}, {Herranz}, {Hildebrandt}, {Hivon}, {Huang}, {Jaffe}, {Jones}, {Karakci}, {Keih{\"a}nen},
  {Keskitalo}, {Kiiveri}, {Kim}, {Kisner}, {Knox}, {Krachmalnicoff}, {Kunz}, {Kurki-Suonio}, {Lagache}, {Lamarre}, {Lasenby}, {Lattanzi}, {Lawrence}, {Le Jeune}, {Lemos}, {Lesgourgues}, {Levrier}, {Lewis}, {Liguori}, {Lilje}, {Lilley}, {Lindholm}, {L{\'o}pez-Caniego}, {Lubin}, {Ma}, {Mac{\'\i}as-P{\'e}rez}, {Maggio}, {Maino}, {Mandolesi}, {Mangilli}, {Marcos-Caballero}, {Maris}, {Martin}, {Martinelli}, {Mart{\'\i}nez-Gonz{\'a}lez}, {Matarrese}, {Mauri}, {McEwen}, {Meinhold}, {Melchiorri}, {Mennella}, {Migliaccio}, {Millea}, {Mitra}, {Miville-Desch{\^e}nes}, {Molinari}, {Montier}, {Morgante}, {Moss}, {Natoli}, {N{\o}rgaard-Nielsen}, {Pagano}, {Paoletti}, {Partridge}, {Patanchon}, {Peiris}, {Perrotta}, {Pettorino}, {Piacentini}, {Polastri}, {Polenta}, {Puget}, {Rachen}, {Reinecke}, {Remazeilles}, {Renzi}, {Rocha}, {Rosset}, {Roudier}, {Rubi{\~n}o-Mart{\'\i}n}, {Ruiz-Granados}, {Salvati}, {Sandri}, {Savelainen}, {Scott}, {Shellard}, {Sirignano}, {Sirri}, {Spencer}, {Sunyaev}, {Suur-Uski}, {Tauber}, {Tavagnacco},
  {Tenti}, {Toffolatti}, {Tomasi}, {Trombetti}, {Valenziano}, {Valiviita}, {Van Tent}, {Vibert}, {Vielva}, {Villa}, {Vittorio}, {Wandelt}, {Wehus}, {White}, {White}, {Zacchei}, \& {Zonca}}]{PlanckCollaboration2020}
{Planck Collaboration}, {Aghanim}, N., {Akrami}, Y., {et~al.} 2020, \aap, 641, A6, \dodoi{10.1051/0004-6361/201833910}

\bibitem[{{Reed} {et~al.}(2007){Reed}, {Bower}, {Frenk}, {Jenkins}, \& {Theuns}}]{Reed2007}
{Reed}, D.~S., {Bower}, R., {Frenk}, C.~S., {Jenkins}, A., \& {Theuns}, T. 2007, \mnras, 374, 2, \dodoi{10.1111/j.1365-2966.2006.11204.x}

\bibitem[{{Regan}(2023)}]{Regan23}
{Regan}, J. 2023, The Open Journal of Astrophysics, 6, 12, \dodoi{10.21105/astro.2210.04899}

\bibitem[{{Sabti} {et~al.}(2022{\natexlab{a}}){Sabti}, {Mu{\~n}oz}, \& {Blas}}]{Sabti2022PRD}
{Sabti}, N., {Mu{\~n}oz}, J.~B., \& {Blas}, D. 2022{\natexlab{a}}, \prd, 105, 043518, \dodoi{10.1103/PhysRevD.105.043518}

\bibitem[{{Sabti} {et~al.}(2022{\natexlab{b}}){Sabti}, {Mu{\~n}oz}, \& {Blas}}]{Sabti2022ApJ}
---. 2022{\natexlab{b}}, \apjl, 928, L20, \dodoi{10.3847/2041-8213/ac5e9c}

\bibitem[{{Sheth} \& {Tormen}(1999)}]{Sheth1999}
{Sheth}, R.~K., \& {Tormen}, G. 1999, \mnras, 308, 119, \dodoi{10.1046/j.1365-8711.1999.02692.x}

\bibitem[{{Springel}(2005)}]{Springel2005}
{Springel}, V. 2005, \mnras, 364, 1105, \dodoi{10.1111/j.1365-2966.2005.09655.x}

\bibitem[{{Tacchella} {et~al.}(2018){Tacchella}, {Bose}, {Conroy}, {Eisenstein}, \& {Johnson}}]{Tacchella2018}
{Tacchella}, S., {Bose}, S., {Conroy}, C., {Eisenstein}, D.~J., \& {Johnson}, B.~D. 2018, \apj, 868, 92, \dodoi{10.3847/1538-4357/aae8e0}

\bibitem[{{Tkachev} {et~al.}(2024){Tkachev}, {Pilipenko}, {Mikheeva}, \& {Lukash}}]{Tkachev2024}
{Tkachev}, M.~V., {Pilipenko}, S.~V., {Mikheeva}, E.~V., \& {Lukash}, V.~N. 2024, \mnras, 527, 1381, \dodoi{10.1093/mnras/stad3279}

\bibitem[{{Visbal} {et~al.}(2015){Visbal}, {Haiman}, \& {Bryan}}]{Visbal15}
{Visbal}, E., {Haiman}, Z., \& {Bryan}, G.~L. 2015, \mnras, 453, 4456, \dodoi{10.1093/mnras/stv1941}

\bibitem[{{Wise} {et~al.}(2019){Wise}, {Regan}, {O'Shea}, {Norman}, {Downes}, \& {Xu}}]{Wise19}
{Wise}, J.~H., {Regan}, J.~A., {O'Shea}, B.~W., {et~al.} 2019, \nat, 566, 85, \dodoi{10.1038/s41586-019-0873-4}

\bibitem[{{Xing} {et~al.}(2023){Xing}, {Zhao}, {Liu}, {Heger}, {Han}, {Aoki}, {Chen}, {Ishigaki}, {Li}, \& {Zhao}}]{Xing2023}
{Xing}, Q.-F., {Zhao}, G., {Liu}, Z.-W., {et~al.} 2023, \nat, 618, 712, \dodoi{10.1038/s41586-023-06028-1}

\bibitem[{{Yeh} {et~al.}(2023){Yeh}, {Smith}, {Kannan}, {Garaldi}, {Vogelsberger}, {Borrow}, {Pakmor}, {Springel}, \& {Hernquist}}]{Yeh23}
{Yeh}, J. Y.~C., {Smith}, A., {Kannan}, R., {et~al.} 2023, \mnras, 520, 2757, \dodoi{10.1093/mnras/stad210}

\end{thebibliography}
\bibliographystyle{aasjournal}

\end{document}